# Testing the assumptions of linear prediction analysis in normal vowels

M. A. LITTLE[a)] *Applied Dynamical Systems Research Group, Oxford Centre for Industrial and Applied Mathematics, Oxford University, United Kingdom*, `littlem@maths.ox.ac.uk`, `http://www.maths.ox.ac.uk/ads`

P. E. McSHARRY *Oxford Centre for Industrial and Applied Mathematics and Engineering Science, Oxford University, United Kingdom*

I. M. MOROZ *Applied Dynamical Systems Research Group, Oxford Centre for Industrial and Applied Mathematics, Oxford University, United Kingdom*

S. J. ROBERTS *Pattern Analysis Research Group, Engineering Science, Oxford University, United Kingdom*

**Abstract**

This paper develops an improved surrogate data test to show experimental evidence, for all the simple vowels of US English, for both male and female speakers, that Gaussian linear prediction analysis, a ubiquitous technique in current speech technologies, cannot be used to extract all the dynamical structure of real speech time series. The test provides robust evidence undermining the validity of these linear techniques, supporting the assumptions of either dynamical nonlinearity and/or non-Gaussianity common to more recent, complex, efforts at dynamical modelling speech time series. However, an additional finding is that the classical assumptions cannot be ruled out entirely, and plausible evidence is given to explain the success of the linear Gaussian theory as a weak approximation to the true, nonlinear/non-Gaussian dynamics. This supports the use of appropriate hybrid linear/nonlinear/non-Gaussian modelling. With a calibrated calculation of statistic and particular choice of experimental protocol, some of the known systematic problems of the method of surrogate data testing are circumvented to obtain results to support the conclusions to a high level of significance.

## I. Introduction

This paper develops an improved method of surrogate data testing (a formal hypothesis test), and thereby demonstrates more reliable experimental evidence that the assumptions of Gaussian linear prediction of speech time series cannot explain all the dynamics of real, normal vowel speech time series. By making a calibrated calculation of the simple non-Gaussian measure of *time-delayed mutual information* (a generalisation of the concept of autocorrelation), while ensuring that the *surrogates* contain no detectable non-Gaussianity, Fig. 1 demonstrates that, for simple, stationary, normal vowel time series of a certain length, the null hypothesis of a Gaussian stochastic process is false to a high level of significance.

The core of most modern, established speech technology is the classical linear theory of speech production (Fant, 1960), bringing together the well-developed subjects of linear digital signal processing and linear acoustics to process and analyse speech time series. The biophysical, acoustic assumption that the vocal tract can be modelled as a linear resonator leads naturally to the use of digital filtering and linear prediction analysis (Markel and Gray, 1976), techniques that are based upon classical statistical signal processing, in turn relying upon a cluster of mathematical results from linear systems theory and Gaussian, ergodic random processes. With these methods, it is possible to separate the vocal tract resonances from the driving force of the vocal folds during voiced sounds such as vowels (this technique is demonstrated in, for example Wong et al. (1979)).

However, the biomechanics of speech cannot be entirely linear (Kubin, 1995). There are several potential sources of nonlinearity in speech. A list of these should include turbulent gas dynamics (Teager and Teager, 1989), but also nonlinear vocal fold dynamics due to the interaction between nonlinear aerodynamics and the vocal folds (Story, 2002), feedback between vocal tract resonances and the vocal folds (Quatieri, 2002) and nonlinear vocal fold tissue properties (Chan, 2003). This list is certainly not exhaustive. Simplified vocal fold models also show hysteresis (Lucero, 1999), and in pathological cases, evidence for nonlinear bifurcations have been observed in experiments on excised larynxes, and nonlinear models have replicated these observations (Herzel et al., 1995).

More recently, there has been growing interest in applying tools from nonlinear and non-Gaussian time series analysis to speech time series attempting to characterise and exploit these nonlinear/non-Gaussian phenomena (Kubin, 1995). Algorithms for finding fractal dimensions (Maragos and Potamianos, 1999) and Lyapunov exponents (Banbrook et al., 1999) have both been applied, giving evidence to support the existence of nonlinearities and possibly chaos in speech. Tools that attempt to capture other nonlinear effects have been developed and applied Maragos *et al.* (2002). Speech time series have been analysed using higher-order statistics such as the bicoherence and bispectrum (Fackrell, 1996), providing evidence against the existence quadratic nonlinearities detectable with third-order statistical moments, although higher-order moments were not investigated. Further, Bayesian Markov chain Monte Carlo methods (Godsill, 1996) have also been used. There have been several attempts to capture nonlinear dynamical structure in speech. Thus, local linear (Mann, 1999), global polynomial (Kubin, 1995), regularised radial basis function (Rank, 2003) and neural network (Wu *et al.*, 1994) methods have all been used to try to build compact models that can regenerate speech time series, with varying degrees of success.

There are however, formidable numerical, theoretical and algorithmic problems associated with the calculation of nonlinear dynamical quantities such as Lyapunov exponents or attractor dimensions for real speech time series, casting doubt over whether a nonlinear description can be justified by the data (McSharry, 2005). Circumventing some of these difficulties, the method of surrogate data testing is a formal hypothesis test allowing an estimate of the likelihood that the null hypothesis of Gaussianity and/or linearity is true.

By this method, Tokuda *et al.* (2001) detected deterministic nonlinear structure in the intercycle dynamics of several Japanese vowels using a specific phase-space nonlinear statistic and spike-and-wave surrogates. Work reported by Miyano *et al.* (2000) used the surrogate data test applied to a Japanese vowel from one male and one female speaker using Fourier surrogates and the same phase-space statistic. In a related work on animal vocalisations, surrogate analysis tests have been carried out using a globally nonlinear versus linear prediction statistic and Fourier surrogates (Tokuda *et al.*, 2002). These studies report significant evidence of deterministic nonlinear structure. Improving the reliability of these results for speech time series is the main aim of the present study, since one of the problems with the surrogate data test is that it is easy to spuriously discount the null hypothesis, as discussed in general by Kugiumtzsis (2001) and McSharry *et al.* (2003). This is because any systematic errors inadvertently introduced while carrying out the test cause a *bias towards rejection* of the null hypothesis.

The organisation of this paper is as follows. Sec. II. motivates the construction of a surrogate data test method with improved reliability, and Sec. III. explains the statistic used. Sec. IV. details the method used to construct the surrogates, and Sec. V. applies the method to an example of a known, nonlinear dynamical system. Sec. VI. then applies the method to normal vowels. Finally, Sec. VII. interprets the results obtained, and Sec. VIII. contains a summary and suggestions for future work.

## II. Problems with Surrogate Data Tests Against Gaussian Linearity in Speech

In the classical linear modelling of speech, as common to most current speech technology (Kleijn and Paliwal, 1995), given an acoustic speech pressure time series, it is typically formally assumed that, over a short interval in time[1], the vocal fold behaviour can be represented as a (short-time) stationary, ergodic, zero-mean, Gaussian random process (Proakis and Manolakis, 1996) that acts a driving input to a linear digital filter, forcing the filter into resonance. Assuming Gaussianity makes it possible to use the efficient *Yule-Walker* equations to find the linear prediction coefficients (Proakis and Manolakis, 1996). In wider circles this amounts to the use of what is known as a linear AR, or autoregressive process.

It is, however, an open question as to whether real speech time series actually do support these assumptions, leading to the need for a test of the following null hypothesis: that the data has been generated by a (short-time) stationary, ergodic, zero-mean, Gaussian random process driving a linear resonator. The desire is to obtain sufficient significance that the null hypothesis can be rejected, achieved by generating an appropriate number of surrogates using the Fourier transform method (Schreiber and Schmitz, 2000). Using this method, the surrogates have the same power spectrum (and thus autocorrelation, by the Wiener-Khintchine theorem (Proakis and Manolakis, 1996)), as the original speech time series, yet have only linear, Gaussian statistical dependencies at different time lags (the specifics of the particular surrogate data analysis technique used are discussed further in the next section).

There follows a discussion of a number of systematic errors that arise with the use of these techniques that motivate the development of a more reliable approach.

Firstly, there are problems with the use of Fourier surrogates due to periodicity artifacts, as used by Miyano *et al.* (2000), introduced because only the finite, *cyclic* autocorrelation is preserved, not the autocorrelation of theoretically infinite duration (Schreiber and Schmitz, 2000). Similarly, for spike-and-wave surrogates, as used by Tokuda *et al.* (2001), the cyclic autocorrelation can differ systematically from the original. Discontinuities can also be introduced if the ends and gradients of each cycle are not matched, so that the high-frequency energy characteristics of the surrogates is different from the original (Small *et al.*, 2001). Therefore, this could lead to false rejection of the null hypothesis because the nonlinear statistic might be sensitive to differences in the cyclic autocorrelation (and hence frequency characteristics), between the original and the surrogates (Kugiumtzsis, 2001). It is at least necessary to discount this possibility.

---

[1] For commercial speech coding standards, typically 20-30ms, which amounts to a maximum of around 480 samples at a sample rate of 16kHz (Kroon and Kleijn, 1995).

Secondly, each additional free parameter in the algorithm used to calculate the nonlinear statistic increases the likelihood that the results of the test will depend upon the choice of these free parameters (Kugiumtzsis, 2001). Since the statistic used to obtain the results in both the cited studies of Miyano *et al.* (2000) and Tokuda *et al.* (2001) requires the choice of several free parameters, varying these may produce a different result, leading again to the spurious rejection of the null hypothesis. Discounting this is also necessary, but for the cited statistic the systematic investigation of such a large parameter space poses formidable challenges.

Thirdly, in the cited studies, the algorithm used to compute the nonlinear statistic is not shown to be insensitive to other, simpler aspects of the time series such as the overall amplitude or mean value (McSharry *et al.*, 2003). Then, for example, if the surrogates all have a different amplitude than the original time series, the nonlinear statistic might reflect this feature as well rather than just the existence of structure consistent or inconsistent with the null hypothesis. This will nearly always lead to the spurious rejection of the null hypothesis, guarding against this possibility is one aim of the present study.

Finally, with the cited studies, analytic values of the statistic are not available for some relevant processes, and so cross-checks of the numerical results with known cases cannot be carried out (for example to ensure that the surrogates really do conform to the null hypothesis).

Circumventing these pitfalls, this study introduces an improved surrogate test, carrying out several precautions in the preparation of the original time series and choice of surrogate generation method, and in the choice, and algorithm, for calculation of the statistic.

## III.  Choice of Statistic

From information theory, a particular metric, the two-dimensional time-delayed mutual information (Fraser and Swinney, 1986) allows the detection of correlations between different time-lagged samples of the time series that cannot be detected by linear statistics based upon second-order moments. Thus rejection of the null hypothesis is confirmation that the time series is generated by a Markov process with non-Gaussian transition probabilities. Note that a specific example of such a process is a purely deterministic nonlinear dynamical system (but rejection of the null hypothesis does not discriminate a deterministic from a stochastic process).

An additional merit to this statistic is the known analytic expression of this value for various linear, Gaussian processes, used to cross-check numerical calculations (Paluš, 1995).

For a function, or time series, $u(t)$, the time-delayed mutual information is:

$$I[u](\tau) = \int_{-\infty}^{\infty}\int_{-\infty}^{\infty} p_{0,\tau}(u,v) \ln p_{0,\tau}(u,v) \mathrm{d}u \mathrm{d}v - 2 \int_{-\infty}^{\infty} p_0(u) \ln p_0(u) \, \mathrm{d}u \qquad (1)$$

and by defining the time delay operator:

$$D^\tau u(t) = u(t+\tau), \qquad (2)$$

then $p_0(u)$ is the probability density of the undelayed time series $D^0 u(t)$, and $p_{0,\tau}(u,v)$ is the joint probability density of $D^0 u(t)$ with a time-delayed copy $D^\tau u(t)$. Equation (1) is a metric of statistical dependence and so is zero only when the samples at time lag zero and $\tau$ are statistically independent, and positive otherwise.

For the null hypothesis of a linear, ergodic, zero-mean Gaussian process, with knowledge of the covariance matrix:

$$\mathbf{C} = \left[\begin{array}{cc} \sigma_{0,0} & \sigma_{\tau,0} \\ \sigma_{0,\tau} & \sigma_{\tau,\tau} \end{array}\right], \qquad (3)$$

where $\sigma_{i,j}$ is the covariance between $D^i u(t)$ and $D^j u(t)$, the analytic value of the mutual information is given by:

$$I_\mathrm{L}[u](\tau) = \frac{1}{2} \ln \sigma_{0,0} + \frac{1}{2} \ln \sigma_{\tau,\tau} - \frac{1}{2} \ln \lambda_1 - \frac{1}{2} \ln \lambda_2, \qquad (4)$$

where $\lambda_1$ and $\lambda_2$ are the eigenvalues of $\mathbf{C}$.

For an independent, identically distributed, zero-mean Gaussian process $e(t)$ of variance $\sigma^2$, the mutual information is:

$$I[e](\tau) = \left\{ \begin{array}{ll} \frac{1}{2}\left[\ln\left(2\pi\sigma^2\right) + 1\right] & \text{if } \tau = 0, \\ 0 & \text{otherwise.} \end{array} \right. \qquad (5)$$

Note that $I_\mathrm{L}[u] = I[e]$ if there are no linear autocorrelations in a time series of variance $\sigma^2$. Typically, the integral expression for $I[u]$ is approximated using summations over discrete probabilities of partitions of $u$ (Kantz and Schreiber, 1997), precluding the direct comparison between the value of this expression with the analytical expression $I_\mathrm{L}[u]$ (Paluš, 1995). However, calculating the integral numerically gives one way that a direct comparison may be achieved, using (first-order Euler) discrete numerical integration:

$$I_{\mathrm{N}}[u](\tau) \approx \sum_{i=1}^{Q} \sum_{j=1}^{Q} p_{0,\tau}(u_i, u_j) \ln p_{0,\tau}(u_i, u_j) \Delta u^2 - 2 \sum_{i=1}^{Q} p_0(u_i) \ln p_0(u_i) \Delta u \qquad (6)$$

with $u_i = i\Delta u + u_{\min}$ and $Q$ the number of uniform discretisation intervals $\Delta u$ that subdivide the full range of the time series[2].

Numerical estimation of the densities $p_0(u_i)$ and $p_{0,\tau}(u_i, u_j)$ can then be carried out by using a quantisation, or box operator:

$$\delta_{\Delta u}(u, w) = \begin{cases} 1 & \text{if } w - \Delta u/2 \leq u < w + \Delta u/2 \\ 0 & \text{otherwise,} \end{cases} \qquad (7)$$

where $w$ is the quantisation centre, and $\Delta u$ is the quantisation interval. Then the estimated densities are:

$$p_0(u_i) = \frac{1}{N \Delta u} \sum_{n=1}^{N} \delta_{\Delta u}\left(D^0 u(n), u_i\right) \qquad (8)$$

and

$$p_{0,\tau}(u_i, u_j) = \frac{1}{(N-\tau)\Delta u^2} \sum_{n=1}^{N-\tau} \delta_{\Delta u}\left(D^0 u(n), u_i\right) \delta_{\Delta u}\left(D^\tau u(n), u_j\right) \qquad (9)$$

Here, $N$ is the length of the time series. In the limit of small $\Delta u$, $I_{\mathrm{N}}[u]$ converges to $I[u]$.

The estimation of the densities $p_0(u)$ and $p_{0,\tau}(u,v)$ from the data is biased due to the finite time series length $N$ and the finite partitions $Q$. Thus $I_{\mathrm{N}}[u]$ systematically overestimates $I[u]$, this overestimate varies with the parameters $N$ and $Q$, as suggested by Paluš (1995). Such variation can be analytically approximated as a series expansion, see for example Herzel and Grosse (1997) and Schürmann (2004). Typical variation of $I_{\mathrm{N}}[u]$ upon $N$ and $Q$ is demonstrated in Fig. 2.

One approach to cancel out the effect of such variation is to use the truncated analytic, series expansions and thus correct for the deviations. However, for large ranges of values of $N$ and $Q$, the truncation leads to further systematic error, therefore the approach taken in this study is to cancel out the variation upon $N$ and $Q$ by calibration with the known analytic case $I[e]$. For any pair of parameters $\{N, Q\}$, calculation of $I_{\mathrm{N}}[e]$ over a large number of realisations of $e(t)$ allows the comparison of this with the analytic expression $I[e]$. Finding a best-fit straight line through the mean numerical value $I_{\mathrm{N}}[e]$ for $0 < \tau < \tau_{\max}$ (with $\tau_{\max}$ being the largest lag of interest, typically 200), and subsequently subtracting this line from the values $I_{\mathrm{N}}[u]$ achieves this calibration. Note that this best-fit straight line is usually not horizontal, there is a small, upwards slope.

The calibrated calculation of $I_{\mathrm{N}}[u]$ now allows application of this and the metric $I_{\mathrm{L}}[u]$ to a time series. If the time series has linear, Gaussian correlations then $I_{\mathrm{N}}[u] \approx I_{\mathrm{L}}[u]$ for all values of $N$ and $Q$, to within the magnitude of the errors introduced by box quantisation, finite duration sample correlation matrix estimates $\mathbf{C}$ and discretisation error due to finite summation approximation of the continuous integrals in $I[u]$. The most important application of this calibrated calculation of $I_{\mathrm{N}}[u]$ is to test that the surrogates, are, to within these expected errors, consistent with the null hypothesis. For surrogate time series $s_u(t)$ of $u(t)$, if $I_{\mathrm{N}}[s_u]$ and $I_{\mathrm{L}}[s_u]$ stay within the expected tolerance given the above errors, this instils confidence that the surrogates actually allow a test against a linear Gaussian process.

## IV. Choice of Surrogate Construction Method

In addition to the particular statistic used, the results of a surrogate data test can depend upon the time series data chosen, the methods used to construct the surrogates, and the interactions between them Kugiumtzsis (2001). Surrogates construction requires consideration of all these factors. In this particular application, and with the chosen statistic, periodicity artifacts must be avoided by ensuring that the end points and gradients of each time series match as closely as possible Schreiber and Schmitz (2000). Otherwise, $I_{\mathrm{L}}[u]$ and $I_{\mathrm{L}}[s_u]$ do not coincide, as their covariance matrices differ systematically.

The Fourier transform method (Schreiber and Schmitz, 2000) is used starting with an initial, randomised shuffle of the time series. Shuffling destroys as much of the original dynamical structure in the time series as possible. Imposition of the desired spectral amplitudes from the original time series forces the same cyclic autocorrelation as the original signal, followed by amplitude adjustment to that of a Gaussian process of zero mean and unit variance.

Note that this is slightly different to the typical practice of constraining the probability density using amplitude adjustment to be the same as that of the original time series. Amplitude adjustment was designed as a palliative measure to circumvent the problem that certain statistics can vary systematically with the overall amplitude, as discussed earlier. In the present study, the nonlinear mutual information calculation is insensitive to the overall amplitude, since the numerical probability densities are estimated over the full scale of the time series.

---

[2]Subsequent reference to $I_{\mathrm{N}}$ as the "nonlinear metric", even though it more accurately describes non-Gaussianity, provides an expressive shorthand.

However, for the present application and choice of statistic, surrogates are required with amplitude distributions constrained to be the same as a Gaussian process, rather than to be the same as the original time series. This is because the test is against the null hypothesis of a linear, Gaussian-driven process, using the nonlinear mutual information metric, and for any Gaussian-driven linear process, the amplitude distribution is also Gaussian (this is a consequence of the fact that any linear combination of Gaussian processes is also Gaussian). The nonlinear mutual information is of course sensitive to the entropy of the distribution of the time series, and the original speech time series have a non-Gaussian distribution (this has been demonstrated empirically using various tests, for example, by the use of higher-order statistical moments (Kubin, 1995)). Therefore, surrogates generated by constraining the amplitude distribution to be the same as the original, non-Gaussian speech time series will be inconsistent with the required null hypothesis of this study, and this inconsistency is then detected by the nonlinear mutual information metric[3].

For this reason, in practice, constraining the amplitude distribution to be the same as the original, non-Gaussian speech time series gives, as predicted by theory, a slight, overall increase to the nonlinear mutual information calculation on the surrogates. Although this is small enough that it does not affect the final results, ensuring that theory and practice are in accord by using amplitude distributions constrained to be Gaussian is of more importance here.

Finally, a finding of this study is no difference to the results with the use of the IAAFT (Iterative Amplitude Adjusted Fourier Transform) method (Schreiber and Schmitz, 2000), and so for the sake of computational simplicity this technique is not used.

## V. Application to a Toy Example

Having described the method, this section demonstrates ruling out the null hypothesis of a linear, stochastic, Gaussian process on a toy, deterministic nonlinear example.

Figure 3(a) shows, plotted against the discrete time index $n = 1, 2 \ldots$ the time series $y(n)$ of an order two, autoregressive process (called an AR(2) process), and Fig. 3(b) the $x$-coordinate time series $x(n)$ of the Lorenz system, a simple, third-order nonlinear differential system, for a set of parameters in the chaotic regime. For these two systems, Fig. 4(a) plots both $I_L$ and $I_N$ for the AR(2) process, demonstrating that the two metrics do indeed agree, showing no significant nonlinearity/non-Gaussianity in this time series. Figure 4(b) plots the same for the Lorenz time series, showing that, after a certain time lag $\tau$, the linear and nonlinear metrics begin to diverge significantly and very quickly. This instils confidence that $I_L$ and $I_N$ behave as expected. Figure 4(a) shows that the accumulated sources of error in the calibrated calculation of $I_N$ amount to a small discrepancy in the value over all time lags, but that, unlike Fig. 4(b), the two values always track each other to within a certain small amount, as noted by Paluš (1995)[4].

In all real-world time series some kind of observation noise must be expected. In order to simulate this, Fig. 5(a) shows the Lorenz time series corrupted by zero mean Gaussian noise of around 30% of the maximum amplitude.

Quantification of the significance of the test is best measured using rank-order statistics, because the form of the distribution of the statistic $I_N$ is unknown Schreiber and Schmitz (2000). Requiring a probability of false rejection of the null hypothesis of $P\%$, generating $M = (0.01P)^{-1} - 1$ surrogates allows the (one-sided) test of the null hypothesis to a significance level of $S = 100\% - P\%$. The probability that $I_N$ is largest on the original time series is $P\%$ as intended. This study sets a significance level of $S = 95\%$, so that $P = 5\%$ and hence $M = 19$ surrogates are generated, one of which is shown in Fig. 5(b).

Although familiarity with the Lorenz system might allow detection of the difference by eye, $x(n)$ and $s_x(n)$ are very similar, and as shown in Fig. 6(a) the linear statistics $I_L[x]$ and $I_L[s_x]$ are practically indistinguishable, and the full extent of variation of $I_L[s_x]$ is very small. Furthermore, Fig. 6(b) shows that the nonlinear metric on the surrogates $I_N[s_x]$ tracks the linear metric on the surrogates to within numerical error. Therefore, the surrogates cannot be separated from the original by the linear metric, and the nonlinear metric on the surrogates agrees with the linear metric on the surrogates. Hence confidence is obtained that only linear statistical dependencies are present in the surrogates. Yet, Fig. 6(c) shows that the nonlinear metric on the original $I_N[x]$ is larger than the value of this statistic on the surrogates, for most time lags $\tau > 10$.

This demonstrates that the test is indeed capable of ruling out the null hypothesis for the chaotic system. There are interesting complications in the details though. For a certain range of low lags (say, $\tau \leq 10$), the results do not warrant confidence in rejecting the null hypothesis, because $I_L$ and $I_N$ on the surrogates differ systematically, noted in Kugiumtzsis (2001).

---

[3]This is mostly because the entropy features significantly in the calculation of the nonlinear mutual information at all time lags, even if the statistical dependencies at other time lags could be jointly Gaussian and therefore completely characterised by the autocorrelation.

[4]There is also a small variation in the linear metric since it depends upon the *sample* covariance matrix estimate from the time series.

Table I: Vowels and codenames used in this study.

| Example | Codename |
|---------|----------|
| f<u>a</u>rther | /aa/ |
| b<u>i</u>rd | /er/ |
| b<u>ea</u>t | /iy/ |
| b<u>i</u>t | /ih/ |
| b<u>a</u>t | /ae/ |
| b<u>e</u>t | /eh/ |
| b<u>oo</u>t | /uw/ |
| p<u>u</u>t | /uh/ |
| p<u>o</u>t | /ao/ |
| b<u>u</u>t | /ah/ |

# VI. Application to Normal Vowels

Having demonstrated that by being selective and avoiding known systematic errors in the use of surrogate data methods, ruling out the null hypothesis where it is indeed known to be false is possible, taking into account the conditions under which the test can be said to be valid. The next step is the application of this method to 20 normal, non-pathological vowel time series from the TIMIT database (Fisher *et al.*, 1986) which have been carefully selected to be as short and stationary as possible. These represent ten different US English sounds from randomly selected male and female speakers, covering all the principal, simple vowels. Diphthongs are avoided since they are considered to be nonstationary in the sense that the vocal tract resonances are changing with time. All the time series are recorded under quiet acoustic conditions with minimal background noise, with 16 bits and sample rate 16kHz. The time series have been normalised to an amplitude range of $\pm 1$. Table I lists the vowels and their codenames, and Table II lists the TIMIT source audio file names and lengths in samples each time series[5]. The time series are therefore all approximately 63ms long. Finally, Figs. 7 and 8 shows plots of all the time series $p(t)$.

Figure 9 picks out one of the time series for closer inspection of the associated surrogates. By eye it is fairly easy to separate the surrogates from the original. However, for this same vowel, Fig. 10 shows that the linear metric is identical on both the original time series and the surrogates, and the surrogates are consistent with the null hypothesis as measured by the nonlinear metric. Therefore the surrogates are consistent with the null hypothesis. Figure 1 shows, again for this same sample vowel, the nonlinear statistic applied to the surrogates and the original speech time series.

Figures 11 and 12 show plots comparing the nonlinear metric (calculated with $Q = 20$) applied to a selection of the original speech time series ($I_N[p]$, thick black line), with the median of the nonlinear metric applied to all corresponding surrogates ($I_N[s_p]$, thin, solid black line). The minimum and maximum values of the nonlinear metric for the surrogates are shown (filled grey area with dotted outlines). Over all the time series and at all time lags $1 \leq \tau \leq 50$, there are only two instances out of $50 \times 20 = 1000$ time lags where the nonlinear metric on the original time series is not the largest value.

# VII. Discussion

Figures 11 and 12 show specific examples of the result that there are an insignificant number of cases where the nonlinear metric on the original is not the maximum value. Simultaneous cross-checks show that the surrogates are indistinguishable using second-order statistics, from the original, and, to within the numerical error associated with the computation of the linear and nonlinear metrics, contain no detectable non-Gaussianity. The broad conclusion is that the linear Gaussian null hypothesis can be rejected for most lags for all the vowels tested.

There are however interesting details that are worth pointing out. It appears that in most cases the nonlinear metric applied to the speech time series follows the broad peaks and troughs of the mutual information in the surrogates, with some obvious exceptions. It is the opinion of the authors that this is an indication that the autocorrelation is, to some extent, broadly indicative of the general statistical dependence between samples at specific time lags. This is perhaps one of the reasons why Gaussian linear prediction is a useful technique – since it can capture a broad picture of the dynamical structure of the vowel sounds. However, clearly linear prediction cannot represent all the structure.

---

[5] Microsoft WAV files of these time series and software to carry out the calibrated surrogate data test are available from the URL http://www.maths.ox.ac.uk/~littlem/surrogates/.

7Table II: Sound file sources and sample lengths of time series.

| Time series code-name | TIMIT file name | Length in samples |
| --- | --- | --- |
| faks0_sx223_aa | TEST/DR1/FAKS0/SX223.WAV | 1187 |
| msjs1_sx369_aa | TEST/DR1/MSJS1/SX369.WAV | 914 |
| fcft0_sa1_er | TEST/DR4/FCFT0/SA1.WAV | 1106 |
| mrws0_si1732_er | TRAIN/DR1/MRWS0/SI1732.WAV | 948 |
| fdac1_si844_iy | TEST/DR1/FDAC1/SI844.WAV | 1143 |
| mreb0_si2005_iy | TEST/DR1/MREB0/SI2005.WAV | 1148 |
| fmaf0_si2089_ih | TEST/DR4/FMAF0/SI2089.WAV | 1023 |
| mbwm0_sa1_ih | TEST/DR3/MBWM0/SA1.WAV | 1151 |
| fjwb1_sa2_ae | TRAIN/DR4/FJWB1/SA2.WAV | 1280 |
| mstf0_sa1_ae | TRAIN/DR4/MSTF0/SA1.WAV | 1053 |
| fdkn0_sx271_eh | TRAIN/DR4/FDKN0/SX271.WAV | 1261 |
| mbml0_si1799_eh | TRAIN/DR7/MBML0/SI1799.WAV | 1213 |
| mdbp0_sx186_uw | TRAIN/DR2/MDBP0/SX186.WAV | 951 |
| fmjb0_si547_uw | TRAIN/DR2/FMJB0/SI547.WAV | 1036 |
| futb0_si1330_uh | TEST/DR5/FUTB0/SI1330.WAV | 1043 |
| mcsh0_sx199_uh | TEST/DR3/MCSH0/SX199.WAV | 1051 |
| fcal1_si773_ao | TEST/DR5/FCAL1/SI773.WAV | 983 |
| mbjk0_si2128_ao | TEST/DR2/MBJK0/SI2128.WAV | 930 |
| fmgd0_sx214_ah | TEST/DR6/FMGD0/SX214.WAV | 971 |
| mdld0_si913_ah | TEST/DR2/MDLD0/SI913.WAV | 997 |

A further observation is that this kind of tracking is apparently absent from the comparison between linear and nonlinear metrics applied to the chaotic Lorenz time series, seen in Fig. 4. Also, as seen in most of the curves in Figs. 11 and 12, for very small lags ($\tau < 4$) the nonlinear and linear metrics mostly coincide and the null hypothesis cannot be comfortably rejected.

One alternative interpretation is that the detected non-Gaussianity is actually spectral nonstationarity in the time series. A deliberate precaution of this study is being careful to select short segments of speech that appear to be as regular as possible. However, an additional check carried out, that of calculating the power spectral densities at the beginning, middle and end of the time series, shows that the spectral differences are very slight. Even so, some of the vowels are more irregular than others. Comparing, for example, `faks0_sx223_aa` against `fjwb1_sa2_ae` in Figs. 7 and 8, the former might be considered more stationary than the latter. However, Fig. 11 shows that, even for the apparently stationary vowel `faks0_sx223_aa`, the level of non-Gaussianity is significant (and in fact, even more so than the case for vowel `fjwb1_sa2_ae`, see Fig. 12). In conclusion, therefore, any slight spectral nonstationarity in the vowel time series that could not be eradicated is not a significant factor in the detection of non-Gaussianity.

## VIII. Concluding Remarks

This paper provides evidence, using an improved surrogate data test method, that the linear Gaussian theory of speech production cannot be the whole explanation for the dynamical structure of simple vowels. It reaches this conclusion by first identifying (Sec. II.) and circumventing (Secs. III.–IV.) certain systematic problems with the use the surrogate data analysis test method that affect other studies. It demonstrates the effectiveness of the method on a known nonlinear time series (Sec. V.). Application to real speech time series demonstrates that the null hypothesis of Gaussian driven linearity is false, to a high level of significance (Sec. VI.).

Although the calibrated calculation of $I_N$ leads to agreement with $I_L$ to within a small discrepancy, more satisfactory would be to avoid the calibration altogether. To this end, a complete theoretical explanation of the systematic divergence of this algorithm from the analytic values would be of value, and perhaps also a more sophisticated calculation of $I_N[u]$ involving adaptive partitioning and the use of higher-order numerical integration.

Careful selection of time series where the linear correlations are stationary over the duration is important, but the duration of approximately 60ms is twice as long as the normal frame size used in speech coders. To provide more incentive for the use of sophisticated nonlinear/non-Gaussian methods in practical speech coding, it might be preferable to test the assumptions over shorter time scales. This study finds that applying the test over such short time scales is problematic for the calculation of $I_N$ since it starts to vary substantially from $I_L$, and confidence is weakened that the surrogates are consistent with the null hypothesis. Improvements suggested in the previous paragraph might enable this test to be carried out.

Although this paper avoids spectral nonstationarity, nonstationarity may be measured by many other quantities, such as running mean, variance and higher-order moments, which could be used as additional checks to assess nonstationarity of non-Gaussian correlations in the speech data Fackrell (1996). This would probably require, however, longer time series, making it harder to find natural speech data for which nonstationarity can be avoided.

It would also be interesting to apply the same technique to other speech time series, for example consonants which are supposed to be turbulent yet still amenable to linear prediction analysis Kubin (1995). Diphthongs are supposed to be generated by nonstationary linear dynamics, but perhaps a nonlinear predictor may model these dynamics more naturally. Application of this test to these vowel sounds might be possible. Pathological speech time series are believed to exhibit signs of chaos, and this method might be adapted to the detection of such complex dynamics. For whispering and shouting, the dynamics might diverge from linearity further.

Finally, these results could have implications in LPAS speech coding standards that rely upon Gaussian codebooks Kroon and Kleijn (1995). Although use of linear prediction analysis increases the Gaussianity of the residual with respect to the original speech signal Kubin (1995), this paper suggests that the residual will never be exactly Gaussian. Improvements to the quality of speech coders may be obtained by appropriate, non-Gaussian codebooks.

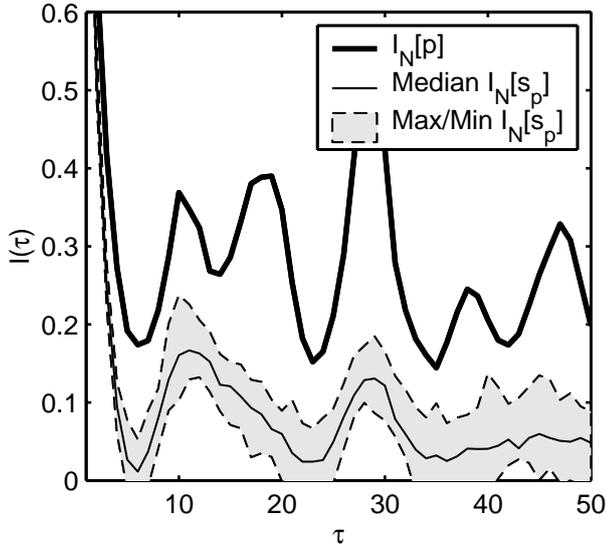

Figure 1: Plot of the nonlinear statistic applied to a particular, simple vowel showing significant discrepancy between the metric applied to linear surrogates (thin black line shows the median value, and the shaded area is bounded by the minimum/maximum values) and the original speech signal (thick black line). The discrepancy demonstrates that to a significance level of 95%, there exists shared dynamical information between samples at most time delays $\tau$ that cannot be accounted for by a purely Gaussian, linear model. (Vowel codename `mbjk0_si2128_ao`, see text).

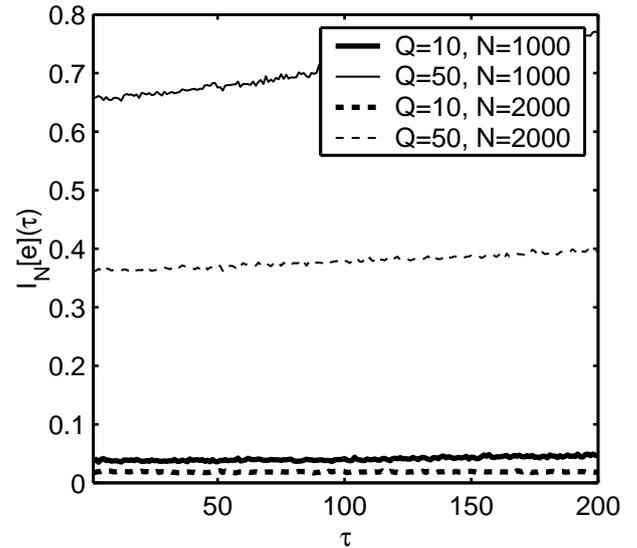

Figure 2: Calibrating the nonlinear mutual information metric on an independent, identically distributed, Gaussian process of the same number of samples $N$ and number of partitions $Q$ of a particular time series. The metric varies systematically with these two parameters, and hence is adjusted to cancel out this parametric dependency.

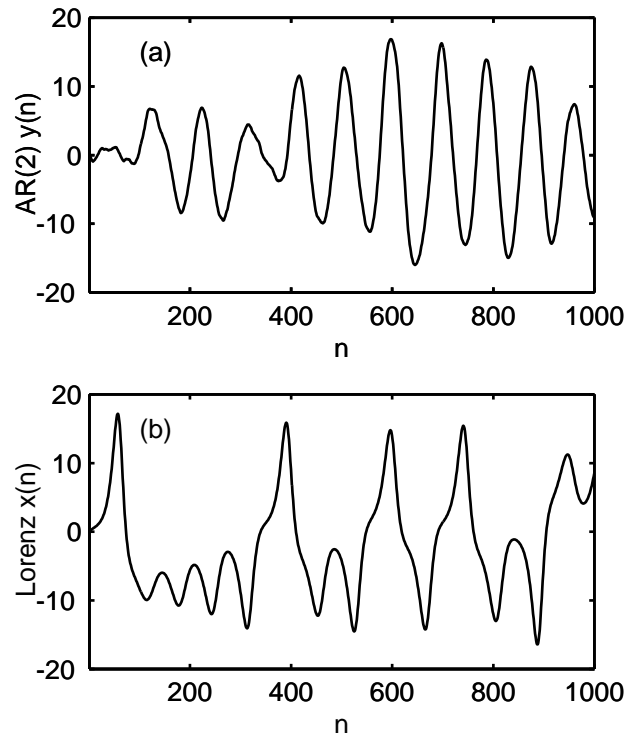

Figure 3: Time series of nonlinear versus linear processes: (a) an order two, Gaussian AR process, (b) the Lorenz system.

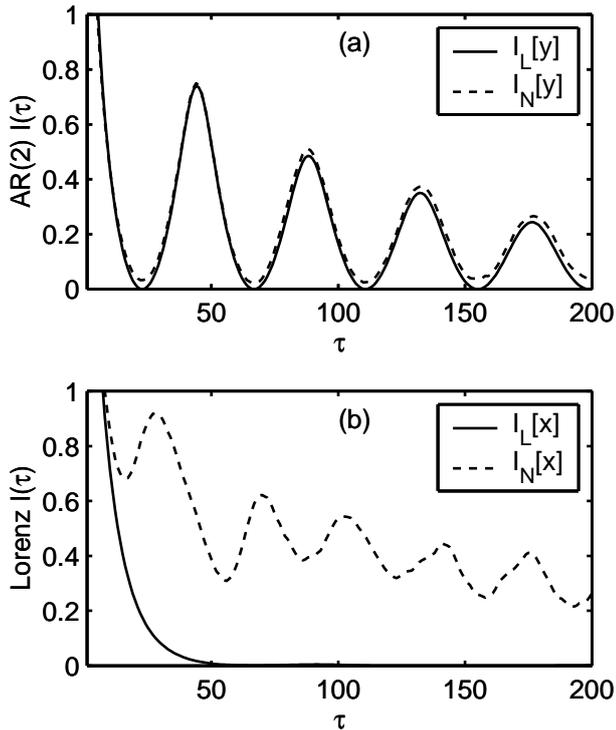

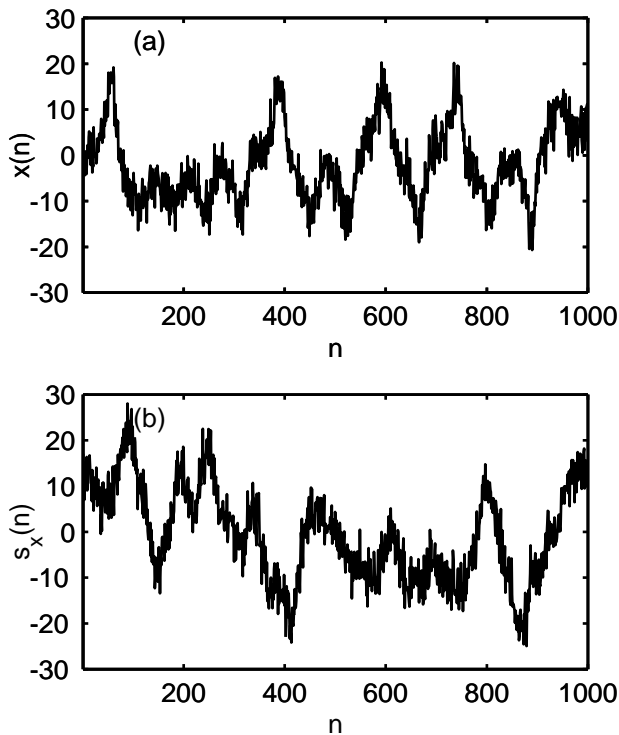

Figure 4: (a) Linear and nonlinear mutual information metrics coincide for a purely linear, Gaussian stochastic process, (b) linear and nonlinear mutual information metrics diverge for a nonlinear, deterministic process. Here $Q = 20$ and $N = 6538$.

Figure 5: (a) Nonlinear Lorenz time series corrupted by Gaussian observation noise, (b) a suitable, linear stochastic surrogate for the above.

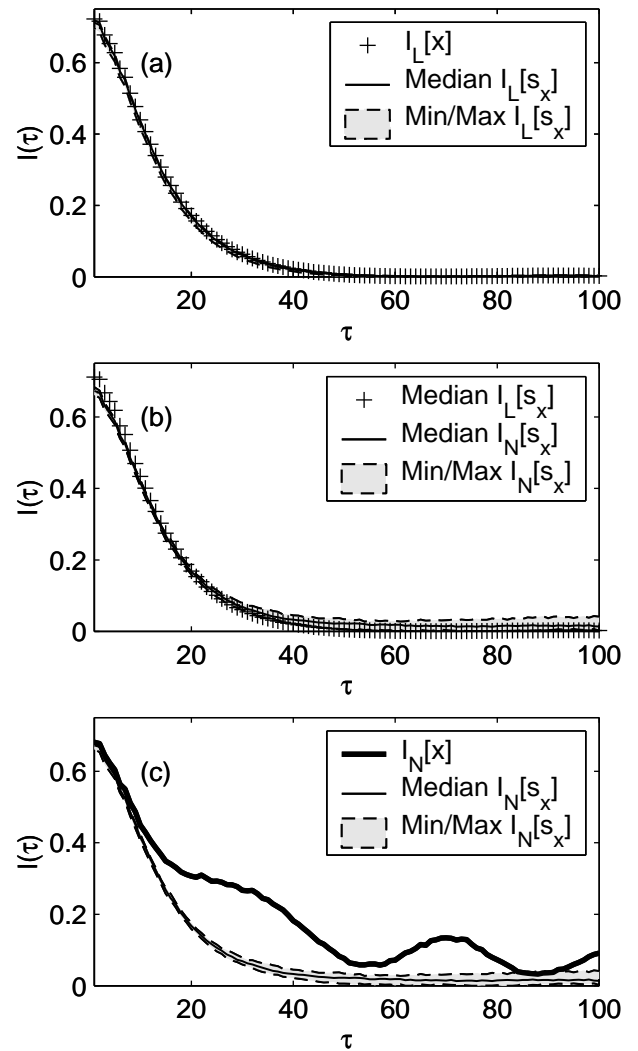

Figure 6: Establishing the difference between the Lorenz system and linear, Gaussian surrogates using cross-application of both linear and nonlinear metrics. (a) Check that surrogates conform to the null hypothesis. Linear mutual information metric on the surrogates has a negligible spread of values, and the median linear metric over all the surrogates coincides with that of the original time series. (b) Check that the nonlinear metric gives the same values as the linear metric for the surrogates which have only linear, Gaussian statistical dependencies. The median of the linear metric on the surrogates tracks the nonlinear metric on the surrogates to within numerical error. (c) Detecting the difference between the deterministic Lorenz time series and the Gaussian linear surrogates. The nonlinear metric on the original time series is the largest value of the nonlinear metric. Combined with the cross-checks in (a) and (b), and given 19 surrogates, the null hypothesis for the Lorenz time series can be ruled out, with a significance level of 95%. Here $Q = 20$ and $N = 6358$.

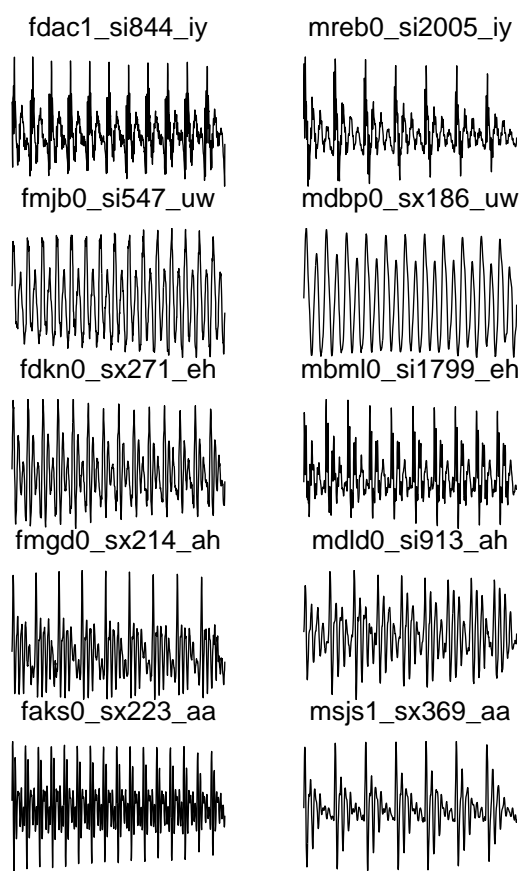

Figure 7: Time series of the first set of vowels used in this study. The codenames are described in Table II. For clarity, the axes labels have been removed, the horizontal axis is time index $n$, the vertical axis is speech pressure $p(n)$.

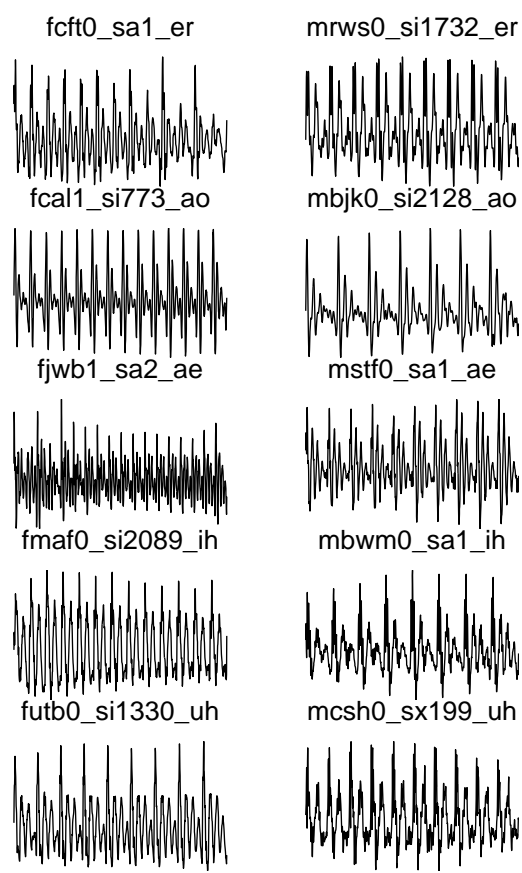

Figure 8: Time series of the second set of vowels used in this study. The codenames are described in Table II. For clarity, the axes labels have been removed, the horizontal axis is time index $n$, the vertical axis is speech pressure $p(n)$.

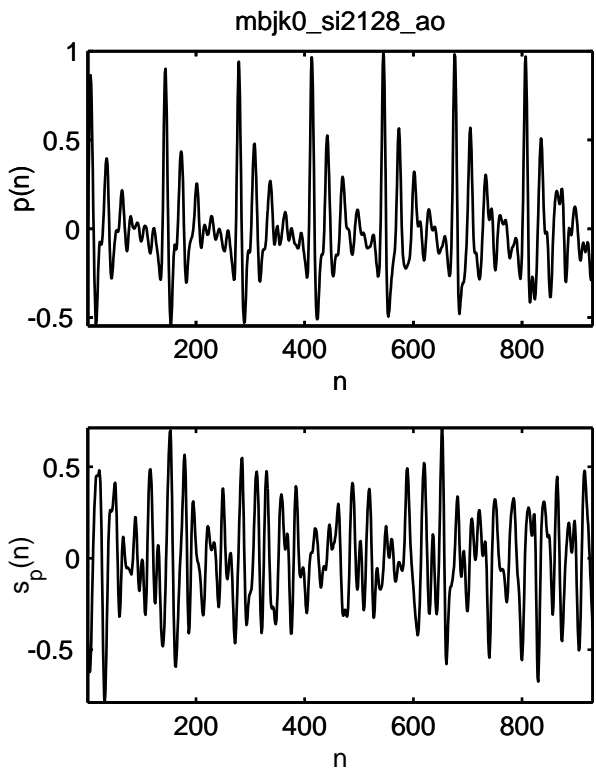

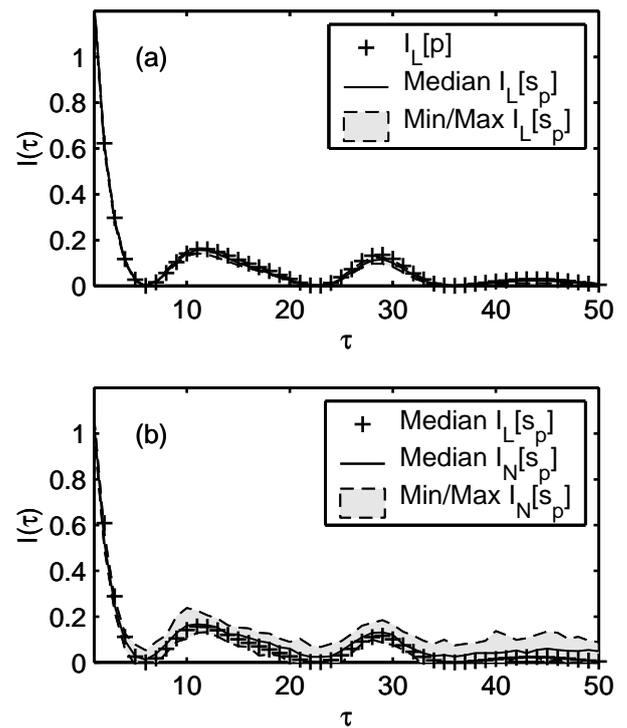

Figure 9: (Top) Time series of a sample vowel from the set, (bottom) one corresponding surrogate.

Figure 10: Surrogate cross-checks for the same sample vowel as selected for Fig. 9. (a) Check that surrogates conform to the null hypothesis. Linear mutual information metric on the surrogates has a negligible spread of values, and the median linear metric over all the surrogates coincides with that of the original time series. (b) Check that the nonlinear metric gives the same values as the linear metric for the surrogates which have only linear, Gaussian statistical dependencies. The median of the linear metric on the surrogates tracks the nonlinear metric on the surrogates to within numerical error.

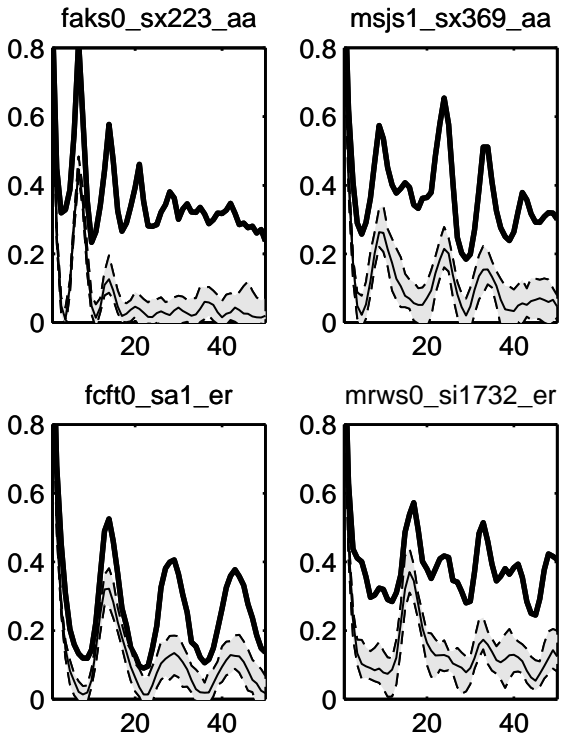

Figure 11: Diverging nonlinear mutual information metrics for simple vowels /aa/ and /er/ and their surrogates. For clarity, the axes labels have been removed, the horizontal axis is time lag $\tau$, the vertical axis is mutual information $I(\tau)$.

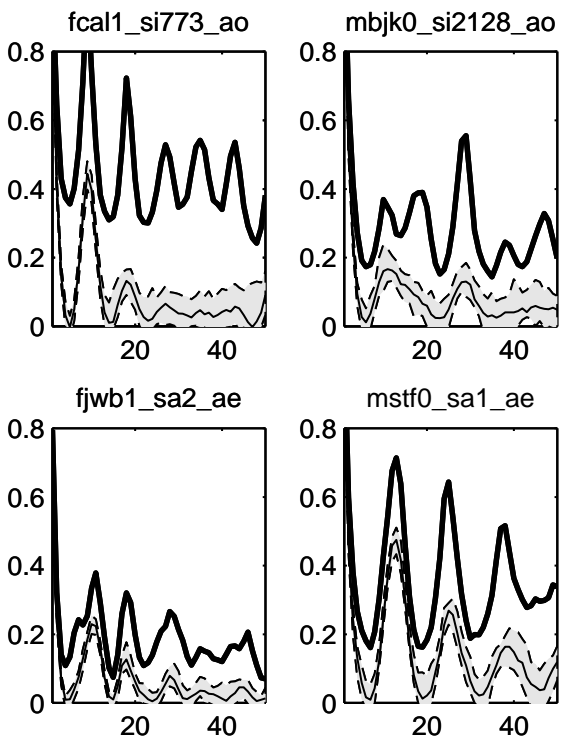

Figure 12: Diverging nonlinear mutual information metrics for simple vowels /ao/ and /ae/ and their surrogates. For clarity, the axes labels have been removed, the horizontal axis is time lag $\tau$, the vertical axis is mutual information $I(\tau)$.

# Acknowledgements

Max Little would like to thank Prof. Stephen McLaughlin for helpful discussions about nonlinear speech processing, and Reason Machete for interesting insights. He acknowledges the EPSRC, UK for financial support.